\documentclass[final]{siamltex}

\usepackage{graphicx}
\usepackage{cite}
\usepackage{color} 
\usepackage{algorithmic}
\usepackage{algorithm}
\usepackage{epsfig}
\usepackage{xspace}
\usepackage{amsfonts}
\usepackage{amsmath}
\usepackage{amssymb}
\usepackage{url}
\usepackage{listings}
\usepackage[labelfont=bf,labelsep=period,justification=raggedright]{caption}
\makeatletter
\renewcommand{\@biblabel}[1]{\quad#1.}
\makeatother
\date{}
\newcommand{\oh}[1]
    {\mbox{$ {\mathcal O}( #1 ) $}}
\newcommand{\eqn}[1]
    {(\ref{eqn:#1})}
\newcommand{\fig}[1]
    {Figure~\ref{fig:#1}}

\newcommand{\sect}[1]
    {Section~\ref{sec:#1}}

\newcommand{\bsf}[1]
    {\textbf{\textsf{#1}}}

\title{Efficient and Scalable Algorithms for Smoothed Particle Hydrodynamics on
    Hybrid Shared/Distributed-Memory Architectures}
\author{Pedro Gonnet\thanks{School of Engineering and Computing Sciences,
    Durham University, Durham, United Kingdom ({\tt pedro.gonnet@durham.ac.uk}).}}

\begin{document}

\lstset{%
    language=C,
    basicstyle=\small\tt,
    numbers=left,
    numberstyle=\tiny
    }

\maketitle

\begin{abstract}
This paper describes a new fast and implicitly parallel
approach to neighbour-finding in multi-resolution
Smoothed Particle Hydrodynamics (SPH) simulations.
This new approach is based on hierarchical cell decompositions and
sorted interactions, within a task-based formulation.
It is shown to be faster than traditional tree-based
codes, and to scale better than domain decomposition-based approaches on
hybrid shared/distributed-memory parallel architectures, e.g. clusters
of multi-cores, achieving a $40\times$ speedup over the Gadget-2
simulation code.
\end{abstract}



\pagestyle{myheadings}
\thispagestyle{plain}
\markboth{P. GONNET}{EFFICIENT AND SCALABLE ALGORITHMS FOR SPH}

\section{Introduction}

Since the past few years, due to the physical limitations
on the speed of individual processor cores, instead of
getting {\em faster}, computers are getting {\em more parallel}.
This increase in parallelism comes mainly in the form of
{\em multi-core} computers, e.g. single computers which
contain more than one computational core sharing a common 
memory bus.
Systems containing up to 64 general-purpose cores are becoming
commonplace, and the number cores can be expected to continue
growing exponentially, e.g.~following Moore's law, much in the
same way processor speeds were up until a few years ago.
This development is not restricted to shared-memory parallel desktop
computers, but also affects modern High-Performance Computing (HPC)
infrastructure which consist mainly of clusters of multi-cores.
Indeed, over the past 5 years, the main factor driving the growth
in cluster performance is the use of shared-memory multi-cores,
and not necessarily an increase in the total number of
nodes/computers used.

For the past 15 years, the predominant paradigm for parallel
computing has been distributed-memory parallelism using MPI
(Message Passing Interface) \cite{ref:Snir1998},
in which large simulations are generally
parallelized by means of data decompositions, i.e.~by assigning
each node or core a portion of the data on which to work.
The cores execute the same code
in parallel, each on its own part of the problem, intermittently exchanging
data with neighbouring cores.
The amount of {\em computation} local to the node is proportional
to the amount of data it contains, e.g.~its {\em volume}, while
the amount of {\em communication} is proportional to the
amount of computation spanning two or more nodes, e.g.~its
{\em surface}.

For very large computations over a moderate number of nodes,
the cost of communication is negligible compared to the
cost of computation, thus providing good parallel efficiency.
However, if the number of nodes increases, or 
smaller problems are considered, the surface-to-volume ratio,
i.e. the ratio of communication to computation,
grows, and the time spent on communication will increasingly
dominate the entire simulation, resulting in a loss of scaling
and parallel efficiency.


Assuming the individual cores do not get any faster,
simulations for which the
maximum degree of parallelism has already been reached will
{\em never} become any faster (see dashed line in \fig{CosmoVolume}).
Ever.
The surface-to-volume ratio problem also means that large
systems which currently parallelize well, if they do not continue
to get larger, will also eventually break down as the number
of cores used for their computation increases.
In order to speed up small simulations, or to continue
scaling for large simulations, new approaches on how
computations are parallelized need to be considered.

In order to address increasingly larger
or more complex problems, high-performance computing software
will need to be able to better exploit the aforementioned increase
in shared-memory parallelism.
Although parallelism and parallel codes are nothing new ---
Many large-scale scientific codes, e.g.~the cosmological
simulation software Gadget-2, can run concurrently
on several thousands of cores ---
the exponential growth of parallelism, and shared-memory
parallelism in particular, provide some interesting 
new challenges.

In this paper, I will describe both a well-known general formulation
for shared memory parallelism, i.e.~task-based parallelism, as well
as its application Smoothed Particle Hydrodynamics simulations.
The task-based approach is extended by the concept of {\em conflicts}
between tasks.
The specific algorithms use several ideas from other particle-based
simulations.
The algorithms are implemented in SWIFT, an Open-Source
platform-independent software for hybrid shared/distributed-memory
parallel simulations which is shown to perform and scale
significantly better than the most popular freely available code
in this area.

\section{Previous work}

In the following, I will give an overview of the underlying equations for
SPH computations, and discuss how they are normally implemented
in multi-resolution simulation codes.

\subsection{Smoothed Particle Hydrodynamics}

Smoothed Particle Hydrodynamics \cite{ref:Gingold1977,ref:Price2012}
 (SPH) uses particles to represent
fluids.
Each particle $p_i$ has a position $\mathbf x_i$,
velocity $\mathbf v_i$, internal energy $u_i$, mass $m_i$,
and a smoothing length $h_i$.
The particles are used to interpolate any quantity $Q$ at any 
point in space as a weighted sum over the particles:
\begin{equation}
    Q(\mathbf r) = \sum_i m_i \frac{Q_i}{\rho_i} W( \|\mathbf r - \mathbf r_i\| , h )
    \label{eqn:interp}
\end{equation}
where $Q_i$ is the quantity at the $i$th particle, $h$ is the
{\em smoothing length}, i.e.~the radius of the sphere within which data will
be considered for the interpolation, and
$W(r,h)$ is the {\em smoothing kernel} or {\em smoothing function}.
Several different forms for $W(r,h)$ exist, each with their own specific
benefits and drawbacks.
In the following, the most common form consisting of a piecewise
cubic polynomial will be used:
\begin{equation*}
    W(r,h) = \frac{8}{\pi h^3} \left\{
        \begin{array}{ll}
            1 - 6\left(\frac{r}{h}\right)^2 + 6\left(\frac{r}{h}\right)^3 & 0 \leq \frac{r}{h} \leq \frac{1}{2}, \\
            2\left( 1 - \frac{r}{h} \right)^3 & \frac{1}{2} < \frac{r}{h} \leq 1 \\
            0 & \frac{r}{h} > 1.
        \end{array}\right.
\end{equation*}

The particle density $\rho_i$ used in \eqn{interp} is itself
computed similarly:
\begin{equation}
    \rho_i = \sum_{r_{ij} < h_i} m_j W(r_{ij},h_i)
    \label{eqn:rho}
\end{equation}
where $r_{ij} = \|\mathbf{r_i}-\mathbf{r_j}\|$ is the Euclidean
distance between particles $p_i$ and $p_j$.
In compressible simulations, the smoothing length $h_i$ of each
particle is chosen such that the weighted number of neighbours
\begin{equation}
    N_{ngb} = \frac{4}{3}\pi h_i^3 \sum_j W( r_{ij} , h_i )
    \label{eqn:nneigh}
\end{equation}
is kept constant to within a given range, e.g.~$\pm 1$.
This can be achieved by applying a Newton iteration to solve
\eqn{nneigh} for $h_i$, where the required derivative
$\partial N_{ngb}/\partial h_i$ is computed alongside \eqn{rho}.

Once the densities $\rho_i$ have been computed,
the time derivatives
of the velocity, internal energy, and smoothing length, which
require $\rho_i$, are computed as followed:
\begin{eqnarray}
    \frac{dv}{dt} & = & -\sum_{r_{ij} < \hat{h}_{ij}} m_j \left[
        \frac{P_i}{\Omega_i\rho_i^2}\nabla_rW(r_{ij},h_i) +
        \frac{P_j}{\Omega_j\rho_j^2}\nabla_rW(r_{ij},h_j) \right], \label{eqn:dvdt} \\ 
    \frac{du}{dt} & = & \frac{P_i}{\Omega_i\rho_i^2} \sum_{r_{ij} < h_i} m_j(\mathbf v_i - \mathbf v_j) \cdot \nabla_rW(r_{ij},h_i), \label{eqn:dudt}
\end{eqnarray}
where $\hat{h}_{ij} = \max\{h_i,h_j\}$, and the particle pressure
$P_i=\rho_i u_i (\gamma-1)$ and correction term
$\Omega_i=1 + \frac{h_i}{3\rho_i}\frac{\partial \rho}{\partial h}$
are computed on the fly.
The polytropic index $\gamma$ is usually set to $\frac{5}{3}$.

The computations in \eqn{rho}, \eqn{dvdt}, and \eqn{dudt}
involve finding all pairs of particles
within range of each other.
Any particle $p_j$ is {\em within range} of a particle $p_i$
if the distance between $p_i$ and $p_j$ is smaller or equal
to the smoothing distance $h_i$ of $p_i$, e.g.~as is done in \eqn{rho}.
Note that since particle smoothing lengths may vary between particles,
this association is not symmetric, i.e. $p_j$ may be in range of
$p_i$, but $p_i$ not in range of $p_j$.
If $r_{ij} < \max\{h_i,h_j\}$, as is required in \eqn{dvdt},
then particles $p_i$ and $p_j$ are within range
{\em of each other}.

The computation thus proceeds in two distinct stages that are
evaluated separately:
\begin{enumerate}
    \item {\em Density} computation: For each particle $p_i$,
        loop over all particles $p_j$ within range of $p_i$ and evaluate
        \eqn{rho}.
    \item {\em Force} computation: For each particle $p_i$,
        loop over all particles $p_j$
        within range of each other and evaluate \eqn{dvdt} and \eqn{dudt}.
\end{enumerate}
The identification of these interacting particle pairs,
as will be shown in the following sections, incurs the main computational
cost, and therefore also presents the main challenge in implementing efficient
SPH simulations.

\subsection{Tree-based approaches}

In its simplest formulation, all particles in an SPH simulation have
a constant smoothing length $h$.
In such a setup, finding the particles in range of any other particle
is similar to Molecular Dynamics simulations, in which all particles
interact within a constant cutoff radius, and approaches which are used
in the latter, e.g. cell-linked lists
\cite{ref:Allen1989} or Verlet lists \cite{ref:Verlet1967}
or more efficient variants thereof \cite{ref:Gonnet2012,ref:Gonnet2013}
can be used.
Both approaches are discussed in the context of SPH simulations
in \cite{ref:Dominguez2011} and \cite{ref:Viccione2008}.

The neighbour-finding problem becomes more interesting, or difficult,
in SPH simulations with variable smoothing lengths, i.e.~in which
each particle has its own smoothing length $h_i$, with ranges spawning
up to several orders of magnitude.
In such cases, e.g. in Astrophysics simulations \cite{ref:Gingold1977},
the above-mentioned approaches cease to work efficiently.
Such codes therefore usually rely on {\em trees}
for neighbour finding \cite{ref:Hernquist1989,ref:Springel2005,ref:Wadsley2004},
i.e.~$k$-d trees \cite{ref:Bentley1975} or octrees \cite{ref:Meagher1982}
are used to decompose the simulation space. 
The particle interactions are then computed by traversing the list of
particles and searching for their neighbours in the tree.

Using such trees, it is in principle trivial to parallelize
the neighbour finding and the actual computation on shared-memory
computers,
e.g.~each thread walks the tree for a different particle,
identifies its neighbours and computes its densities and/or
the second derivatives of the physical quantities of interest for 
the time integration.

Despite its simple and elegant formulation, the tree-based
approach to neighbour-finding has three main problems:
\begin{itemize}
    \item Computational efficiency: The cost of finding all neighbours
        of any given particle in the tree is, on average, in \oh{\log N},
        and has worst-case behavior in \oh{N^{2/3}} \cite{ref:Lee1977},
        i.e.~in any case, the computational cost per particle grows with the
        total number of particles $N$.
    \item Cache efficiency: When searching for the neighbours of a
        given particle, the data of all potential neighbours, which may
        not be contiguous in memory, is traversed.
        This leads to scattered memory access patterns that may be
        cache-inefficient. Furthermore, this operation is performed for
        each particle separately, further reducing the chances
        of cache re-use.
        On shared-memory parallel architectures, this problem is of
        particular concern as parts of the cache hierarchy and the
        memory bandwidth are shared between cores, effectively
        reducing both in parallel computations.
    \item Symmetry: The parallel tree search can not exploit symmetry,
        i.e.~a pair $p_i$ and $p_j$ will always be found twice,
        once when walking the tree for each particle. It would, however,
        be sufficient to find it once and update both particles, as most
        of the particle interactions are symmetric.
        If this is done in a shared-memory parallel setup, special
        care muss be taken to avoid concurrency problems when
        two threads update the same particle's data.
\end{itemize}
    
These problems are all inherently linked to the use of
spatial trees, and more specifically their traversal,
for neighbour-finding.

\subsection{Task-based parallelism}

\begin{figure}
    \centerline{\epsfig{file=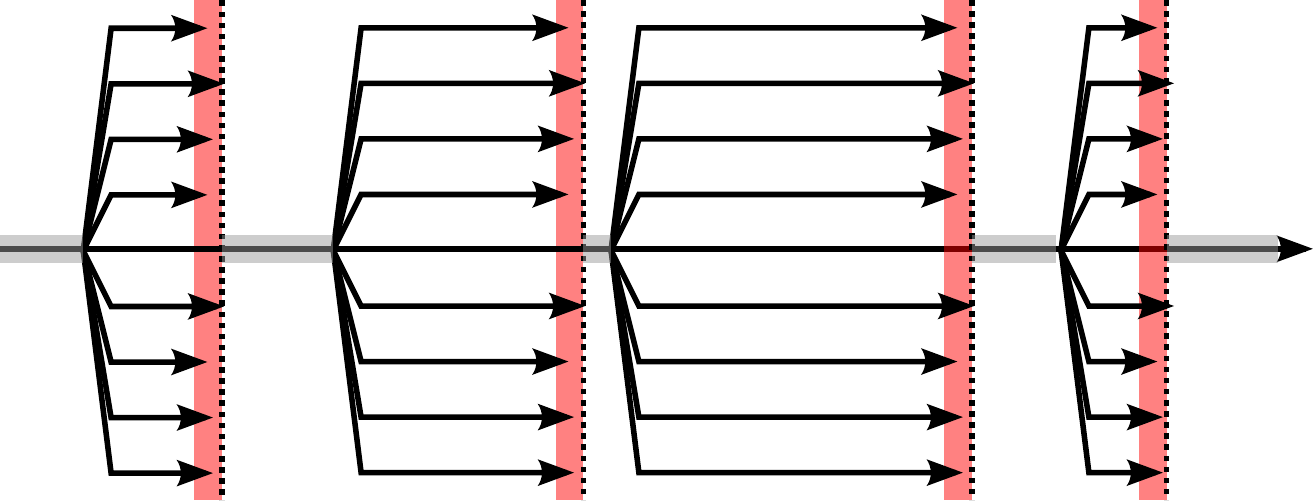,width=0.7\textwidth}}
    
    \caption{Branch-and-bound parallelism as is commonly used in OpenMP.
        The horizontal arrows indicate the program flow over time, and
        branching arrows indicate a parallel section. The dotted vertical
        bars are the synchronization points at the end of each such section.
        Parallel efficiency is lost to two factors: The grey shaded areas
        along the main horizontal area indicate parts of the program that
        do not execute in parallel and restrict the maximum parallel
        efficiency, e.g.~as described by Amdahl's law, and the red
        areas indicate the difference between the fastest and slowest
        threads in each parallel block, i.e.~the time lost to
        individual thread load imbalances and synchronization.
        }
    \label{fig:OMPScaling}
\end{figure}

The arguably most well-known paradigm for shared-memory,
or multi-threaded parallelism is OpenMP \cite{ref:Dagum1998},
in which compiler annotations are used to describe if and when
specific loops or portions of the code can be executed
in parallel.
When such a parallel section, e.g.~a parallel loop, is
encountered, the sections of the loop are split statically
or dynamically over the available threads, each executing
on a single core.
Once all the threads have terminated, the program continues
executing in a single thread.
Unfortunately, this can lead to a lot of inefficient
branch-and-bound
type operations, which generally lead to low performance and
bad scaling on even moderate numbers of cores (see \fig{OMPScaling}).

An additional complication is that this form of shared-memory
parallelism provides no implicit mechanism to avoid or handle
concurrency problems,
e.g.~two threads attempting to modify the same data at the same time,
or data dependencies between them.
These must be implemented explicitly using either redundancy, barriers,
critical sections, or atomic memory operations, which can further degrade
parallel performance.

In order to better exploit shared-memory parallelism, 
a different paradigm is needed, i.e. instead
of annotating an essentially serial computation with parallel
bits, it is preferable to describe the entire computation in a way that
is inherently parallelizable.
One such approach is {\em task-based parallelism}, in which the
computation is divided into a number of computational tasks, which are
then dynamically allocated to a number of processors.
In order to ensure that the tasks are executed in the right
order, e.g.~that data needed by one task is only used once it
has been produced by another task, and that no two tasks
update the same data at the same time, {\em dependencies} between
tasks are specified and strictly enforced by a task scheduler.

Such a set of tasks and dependencies form a Directed Acyclic Graph (DAG),
which the processors can traverse in topological order, picking up and
executing tasks when they have no unresolved dependencies.
Whenever a processor is done with a task, it goes back to
the DAG and looks for a new one or waits until a task
becomes available, until all tasks have been completed.

\begin{figure}
    \centerline{\epsfig{file=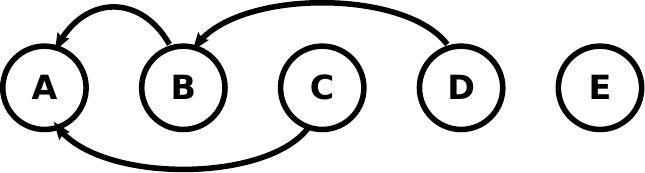,width=0.5\textwidth}}
    
    \caption{Five tasks, A, B, C, D, and E, with their dependencies
        shown as arrows, i.e.~tasks B and C both depend on task A.}
    \label{fig:TasksExample}
\end{figure}

\fig{TasksExample} shows five tasks, A, B, C, D, and E, drawn
as circles, and their dependencies, depicted as arrows.
In this example, tasks B and C depend on task A, and task D
depends on task B.
In a task-based parallel environment, tasks A and E could
be executed first, as they have no unresolved dependencies.
Once task A has been executed, tasks B and C become available.
Task D can only be executed once task B has completed.

Several middle-wares providing such task-based
parallelism exist, e.g.~Cilk \cite{ref:Blumofe1995}, QUARK \cite{ref:QUARK},
StarPU \cite{ref:Augonnet2011}, SMP~Superscalar \cite{ref:SMPSuperscalar},
OpenMP~3.0 \cite{ref:Duran2009}, and Intel's TBB \cite{ref:Reinders2007}.
Cilk and StarPU are implemented as extensions to the C programming
language, whereas SMP Superscalar and OpenMP 3.0 use so-called {\tt pragma}s
to define functions or sections of code that form tasks.
QUARK and Intel's TBB are implemented as compiler-independent
libraries which provide functionality for creating and executing
tasks.

The main advantages of using a task-based approach are
\begin{itemize}
    \item The order in which the tasks are processed is completely
        dynamic and adapts automatically to load imbalances.
    \item If the dependencies and conflicts are specified correctly,
        there is no need for expensive explicit locking, synchronization,
        or atomic operations to deal with most concurrency problems.
    \item Each task has exclusive access to the data it is working on,
        thus improving cache locality and efficiency.
        If each task operates exclusively on a restricted part of the
        problem data, this can lead to hich cache locality and efficiency.
\end{itemize}
Despite these advantages, task-based parallelism is only rarely
used in scientific codes, with the notable exception of
the PLASMA project \cite{ref:Agullo2009}, which is the driving
force behind the QUARK library, and the {\tt deal.II} project
\cite{ref:Bangerth2007} which uses Intel's TBB.
This is most probably due to the fact that, in order to profit
from the many advantages of task-based programming,
for most non-trivial problems, the underlying algorithms must
be redesigned from the bottom up in a task-based way.

\section{Algorithms}

In the following, I will describe both an extension to the
traditional task-based parallel programming model, as well
as a task-based formulation for neighbour-finding and 
particle interactions in SPH simulation.

\subsection{Task-based parallelism with conflicts}

\begin{figure}
    \centerline{\epsfig{file=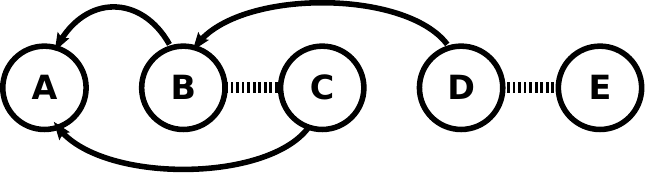,width=0.5\textwidth}}
    
    \caption{Five tasks, A, B, C, D, and E, with their dependencies
        shown as arrows, and their conflicts shown as dashed lines,
        i.e.~tasks B and C, and tasks D and E conflict with each
        other.}
    \label{fig:TasksExampleConflicts}
\end{figure}

I will differ from previous approaches to task-based parallelism
in introducing the concept of {\em conflicts} between tasks.
Conflicts occur when two tasks operate on the same data, 
but the order in which these operations must occur is not defined.
\fig{TasksExampleConflicts} extends the example in the
previous section with conflicts between tasks B and C, and tasks D and E.
In a parallel setup, once task A has been completed, if one processor
picks up task B, then no other processor is allowed to execute
task C until task B has completed, or vice versa.

In previous task-based models, conflicts can be modeled by adding
dependencies between conflicting tasks,
yet this introduces an artificial ordering between the tasks
and imposes unnecessary constraints on the task scheduler
(e.g.~mutual and non-mutual interactions in \cite{ref:Ltaief2012}).

In the following, conflicts are modelled using exclusive {\em locks} on shared
resources, i.e.~a task operating on potentially shared data will
only be scheduled if the executing thread can obtain an exclusive
lock on that data,
thus preventing other tasks using said data to be scheduled concurrently.
This locking mechanism will be described in more detail further on.

\subsection{Spatial decomposition}

Besides the problems described in the previous section,
spatial trees also have the disadvantage that they do not
lend themselves particularly well for task-based computations.
Therefore, in the following, the particle interactions will be described in
terms of {\em hierarchical cell lists}, and the operations thereon.

If $h_\mathsf{max} := \max_i  h_i $ is the maximum smoothing
length of any particle in the simulation,  the simulation domain
is split into rectangular cells of edge length
larger or equal to $h_\mathsf{max}$.

Given such an initial decomposition,
a list of cell {\em self-interactions}, which contains all
non-empty cells in the grid, is generated.
This list of interactions is then extended by the cell
{\em pair-interactions}, i.e. a list of all non-empty cell pairs
sharing either a face, edge, or corner.
For periodic domains, cell pair-interactions must also be
specified for cell neighbouring each other across
periodic boundaries.

In this first coarse decomposition, if a particle $p_j$
is within range of a particle $p_i$, both will be either
in the same cell, or in neighbouring cells for which a
cell self-interaction or cell pair-interaction has been
specified respectively.
These self- and pair-interactions therefore encode,
conceptually at least,
the evaluation of the interactions between all particles
in the same cell, or all interactions between particle pairs
spanning a pair of cells, respectively, i.e.~if the list of
self-interactions and pair-interactions are traversed,
computing all the interactions within each cell and between
each cell pair, respectively, the all the required particle
interactions will have been computed.

In the best case, i.e.~if each cell contains
only particles with smoothing length equal to the cell
edge length, if for any particle $p_i$ each
particle $p_j$ in the same and neighbouring cells is
inspected, only roughly 16\% of the $p_j$
will actually be within range of $p_i$ \cite{ref:Gonnet2007}.
If the cells contain particles who's smoothing length
is less than the cell edge length, this ratio only
gets worse.
It therefore makes sense to refine the cell decomposition
recursively, bisecting each cell along
all spatial dimension whenever (a) the cell contains more than
some minimal number of particles, and (b) the smoothing
length of a reasonable fraction of the particles within
the cell is less than half the cell edge length.
A cell will be referred to as {\em split} if it
has been divided into {\em sub-cells}.

\begin{figure}
    \centerline{\epsfig{file=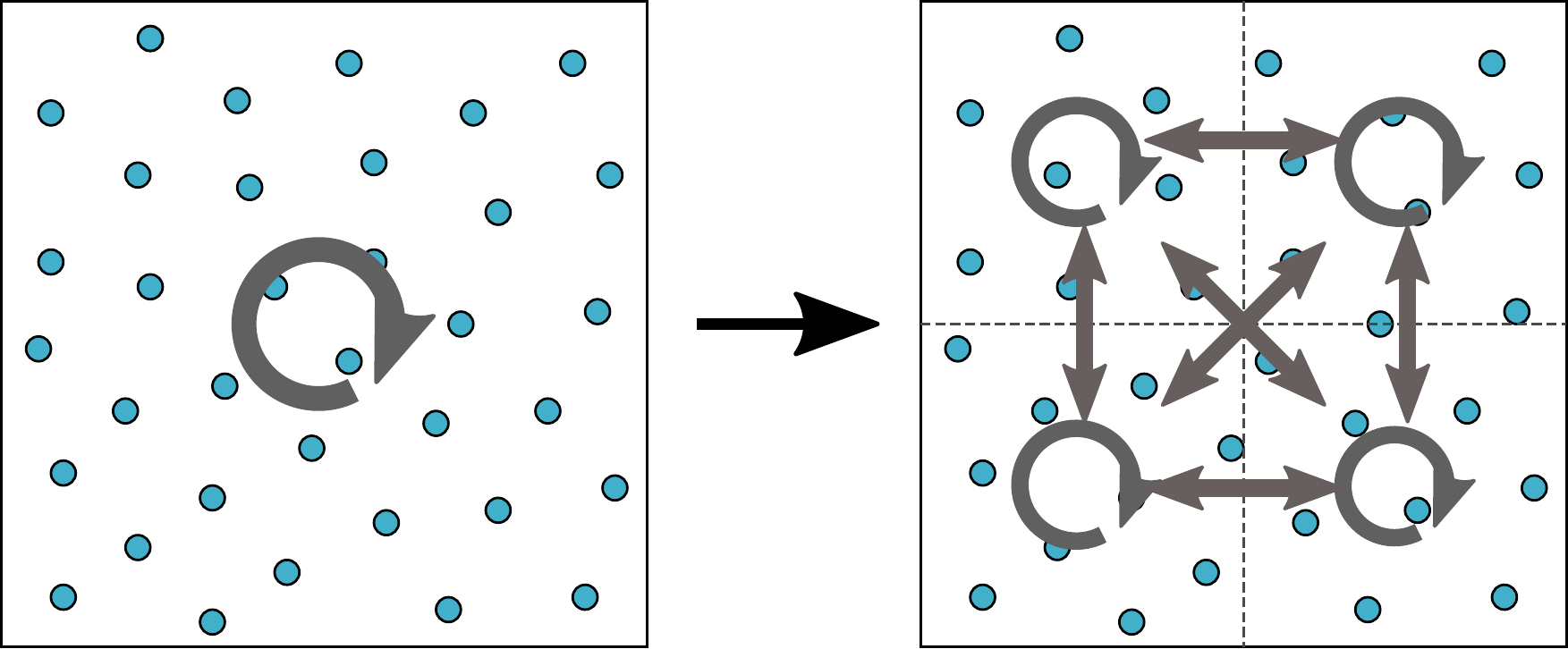,height=0.2\textwidth}}
    
    \caption{Large cells can be split, and their
        self-interaction replaced by the self- and pair-interactions
        of their sub-cells.}
    \label{fig:SplitCell}
\end{figure}

\begin{figure}
    \centerline{\epsfig{file=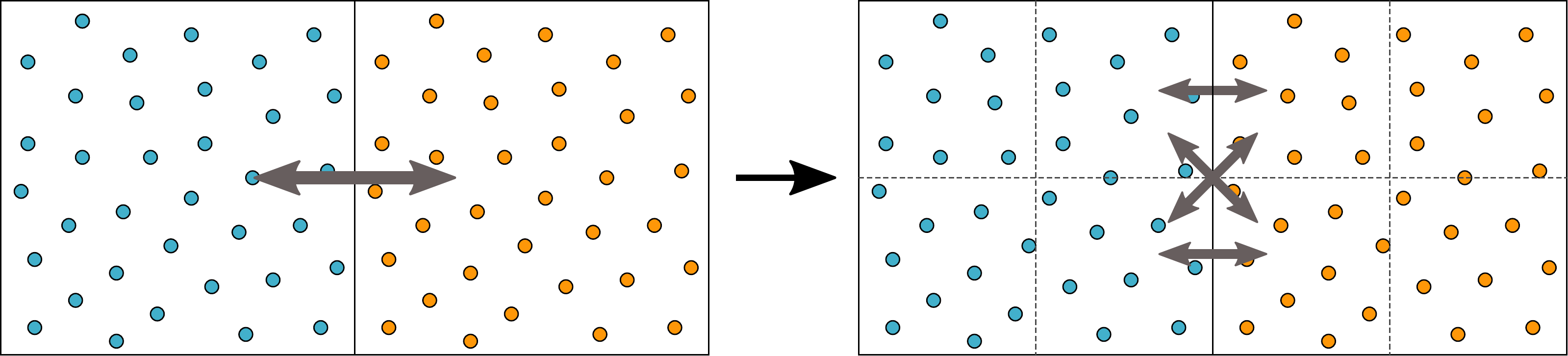,height=0.2\textwidth}}
    
    \caption{If all particles in a pair of interacting cells have a smoothing
        length less or equal to half of the cell edge length, both cells can be
        split, and their pair-interaction replaced by the pair-interactions
        of the neighbouring sub-cells across the interface.}
    \label{fig:SplitPair}
\end{figure}

After the cells have been split, the cell self-interactions
of each cell can be split up into the self-interaction
of its sub-cells and the pair-interactions between
them (see \fig{SplitCell}).
Likewise, the cell pair-interactions between two cells
that have been split can themselves be split up into
the pair-interactions of the sub-cells spanning the
original pair boundary (see \fig{SplitPair}) if, and only if,
all particles in both cells have a smoothing length of
less than half the cell edge length.
If only one cell within a cell pair-interaction has been
split, then the cell pair-interaction is preserved, i.e. the
interactions between the particles in both cells are computed,
yet the split cell is considered as a whole.

If the cells, self-interactions, and pair-interactions are split
in such a way, if two particles are within range of each other,
they will (a) either share a cell for which a cell self-interaction
is defined, or (b) they will be located in two cells which share
a cell pair-interaction.
In order to identify all the particles within range of each other,
it is therefore sufficient to traverse the list of
self-interactions and pair-interactions, and to compute the
interactions therein.

\begin{figure}
    \centerline{\epsfig{file=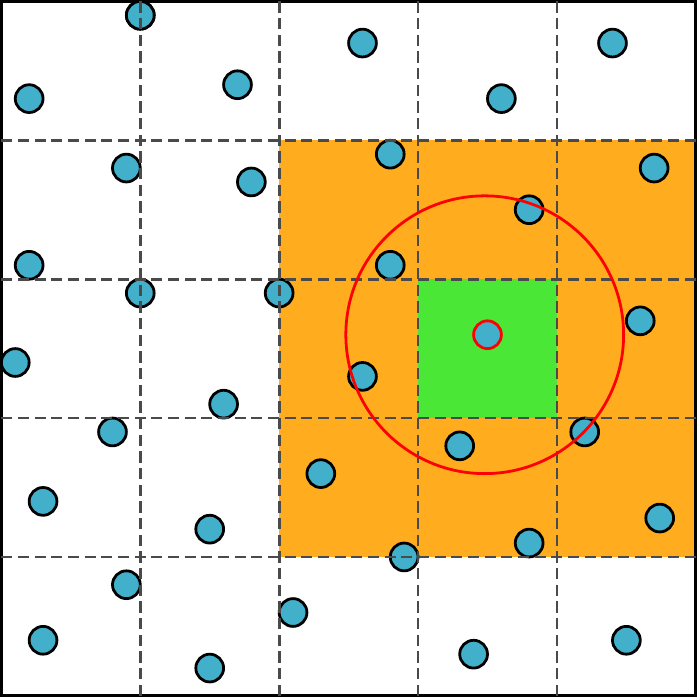,height=0.3\textwidth}}
    
    \caption{Initial spatial decomposition: The space is divided into cells of
        edge length greater or equal to the largest smoothing length in the
        system. All neighbours of any given particle (small red circle) within
        that particle's smoothing length (large red circle) are guaranteed to lie
        either within that particle's own cell (green) or the directly
        adjacent cells (orange).}
    \label{fig:InitialDecomp}
\end{figure}

\subsection{Particle interactions}

The interactions between all particles within the same cell,
i.e. the cell's self-interaction, can be computed by means of
a double {\tt for}-loop over the cell's particle array.
The algorithm, in C-like pseudo code, can be written as follows:

\begin{center}\begin{minipage}{0.8\textwidth}
    \begin{lstlisting}
for (i = 0; i < count - 1; i++) {
  for (j = i + 1; j < count; j++) {
    rij = || parts[i] - parts[j] ||.
    if (rij < h[i] || rij < h[j]) {
      compute interaction.
    }
  }
}
    \end{lstlisting}
\end{minipage}\end{center}

\noindent where {\tt count} is the number of particles in the
cell and {\tt parts} and {\tt h} refers to an array of those
particles' positions and their smoothing lengths respectively.

The interactions between all particles in a pair of cells
can be computed similarly, e.g.:
   
\begin{center}\begin{minipage}{0.8\textwidth}
    \begin{lstlisting}
for (i = 0; i < count_i; i++) {
  for (j = 0; j < count_j; j++) {
    rij = || parts_i[i] - parts_j[j] ||.
    if (rij < h_i[i] || rij < h_j[j]) {
      compute interaction.
    }
  }
}
    \end{lstlisting}
\end{minipage}\end{center}

\noindent where {\tt count\_i} and {\tt count\_j} refer to
the number of particles in each cell and {\tt parts\_i} and
{\tt parts\_j}, and {\tt h\_i} and {\tt h\_j}, refer to the
particles of each cell and their smoothing lengths respectively.

\begin{figure}
    \centerline{\epsfig{file=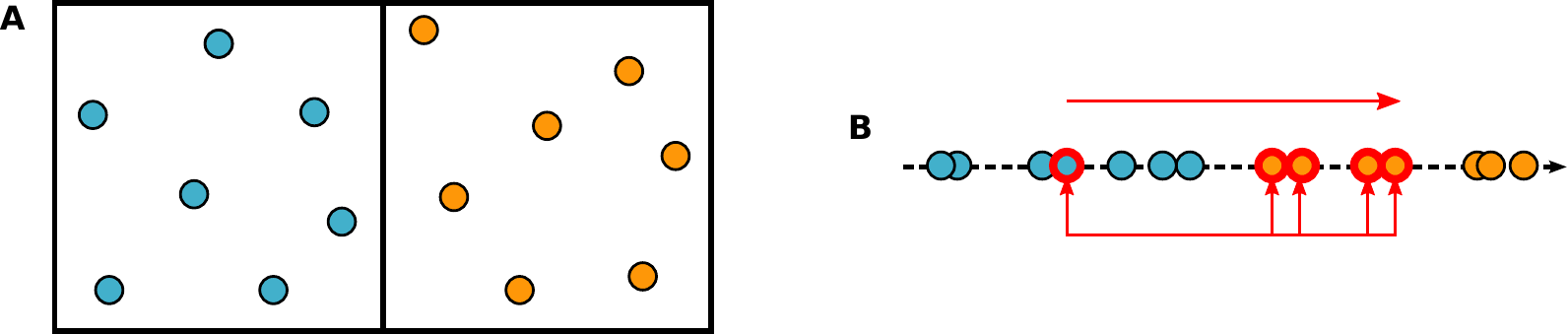,width=0.7\textwidth}}
    
    \caption{Sorted cell pair-interactions. ({\bsf A}) Starting from a pair of
        neighbouring cells, ({\bsf B}) the particles from both cells
        are projected onto the axis joining the centers of the two cells.
        The particles on the left (blue) and right (orange) are
        then sorted in descending and ascending order respectively.
        Each particle on the left is then only interacted with
        the particles on the right within the cutoff radius along the cell axis.
        }
    \label{fig:SortedInteractions}
\end{figure}

As described in \cite{ref:Gonnet2007}, though, using this
naive double {\tt for}-loop, only roughly $33.5\%$, $16.2\%$,
and $3.6\%$ of all particle
pairs between cells sharing a common face, edge, or corner, respectively,
will be within range of each other, leading
to an excessive number of spurious pairwise distance evaluations (line~3).
The sorted cell interactions described therein will be used in order to
avoid this problem, yet with some minor modifications, as
the original algorithm is designed for systems in which the
smoothing lengths of all particles are equal:
The particles in both cells are first sorted along the vector joining
the centers of the two cells, then the
parts $p_i$ on the left are interacted with the sorted parts $p_j$
on the right which are within $h_i$ {\em along the cell pair axis}.
The same procedure is repeated for each particle $p_j$ on the
right, interacting with each other particle $p_i$ on the
left, which is within $h_j$, {\em but not within} $h_i$, along
the cell pair axis (see \fig{SortedInteractions}).
The resulting algorithm, in C-like pseudo-code, can be written as follows:
        
\begin{center}\begin{minipage}{0.8\textwidth}
    \begin{lstlisting}
r_i = parts_i projected onto the cell pair axis
r_j = parts_j projected onto the cell pair axis
ind_i = indices of parts_i sorted w.r.t. r_i in ascending order
ind_j = indices of parts_j sorted w.r.t. r_j in ascending order
for (i = 0; i < count_i; i++) {
  for (jj = 0; jj < count_j; jj++) {
    j = ind_j[jj];
    if (r_i[i] + h_i[i] < r_j[j]) break;
    rij = || parts_i[i] - parts_j[j] ||.
    if (rij < h_i[i]) {
      compute interaction.
    }
  }
}
for (j = 0; j < count_j; j++) {
  for (ii = count_i - 1; ii >= 0; ii--) {
    i = ind_i[i];
    if (r_i[i] < r_j[j] - h_j[j]) break;
    rij = || parts_i[i] - parts_j[j] ||.
    if (rij < h_j[j] && rij > h_i[i]) {
      compute interaction.
    }
  }
}
    \end{lstlisting}
\end{minipage}\end{center}
        
\noindent where {\tt r\_i} and {\tt r\_j} contains the position
the particles of both cells along the cell axis, and
{\tt ind\_i} and {\tt ind\_j} contain the particle indices
sorted with respect to these positions respectively.
The {\tt if}-statements in lines 9 and 19 are needed to check
if there are any particles left within range along the cell
pair axis, and to prematurely exit the innermost loop if
this is not the case.

The particles need to be traversed twice: once to identify
all particles in range of the particles on the left (lines~5--14), and
once to identify all particles in range of the particles on the
right (lines~15--24), but that were not identified in the first loop, 
thus the condition in line~20.
Instead of sorting the particles every time the
pairwise interactions between two cells are computed,
the sorted indices along the 26 possible cell-pair axes can be
pre-computed and stored for each cell.
These sorted indices are, however, symmetric: e.g. the indices
computed for a cell interacting with a cell to its left along the
$x$-axis are the inverse of the indices required for interacting 
with the cell on its right.
It is therefore only necessary to sort 13 sets of indices, and flip
the cells in a cell pair-interaction around when the order
required is the opposite of the order stored, i.e. as is
done in \cite{ref:Gonnet2013}.

\begin{figure}
    \centerline{\epsfig{file=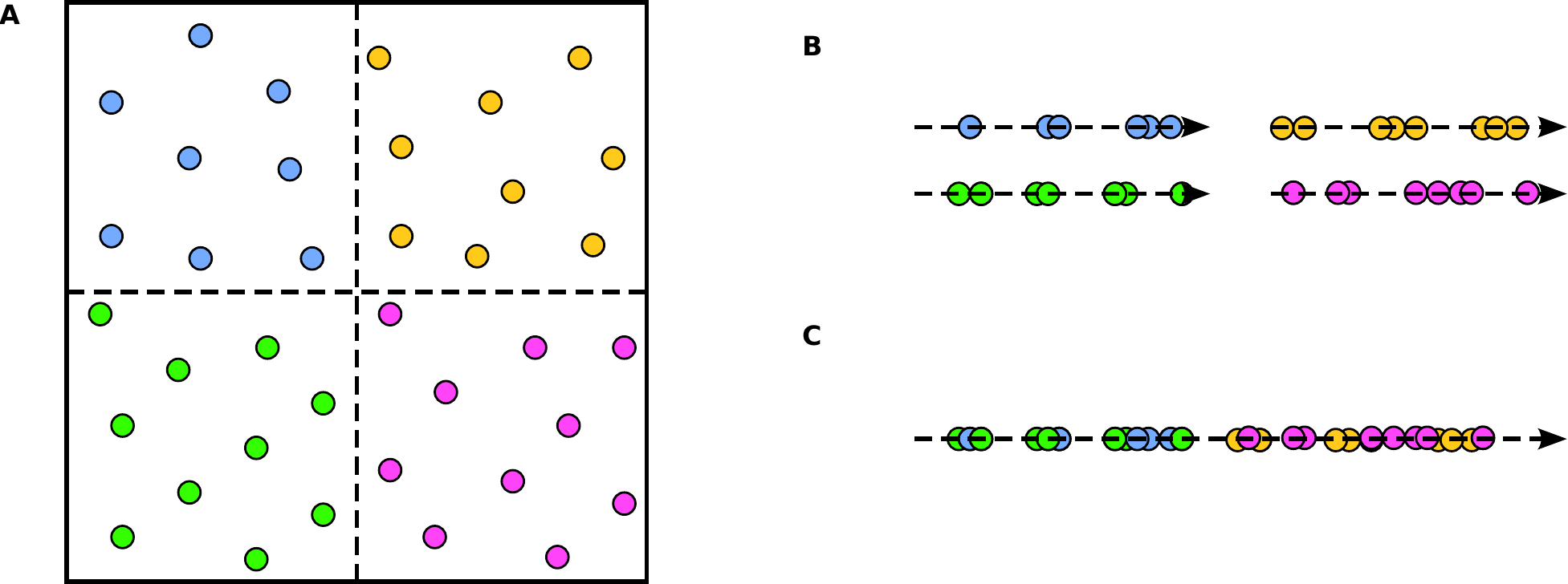,width=0.7\textwidth}}
    
    \caption{Hierarchical cell sorting of ({\bsf A}) a split cell.
        ({\bsf B}) The sub-cells are first sorted individually and
        ({\bsf C}) shifted and merged to produce the sorted list
        of the parent cell.
        }
    \label{fig:HierarchySorting}
\end{figure}

This may still seem like quite a bit of sorting, especially
for the larger, higher-level cells in the simulation.
If, however, a cell is split and its sub-cells have been sorted,
the sorted indices of the higher-level cells can be constructed
by shifting and merging the indices of its eight sub-cells
(see \fig{HierarchySorting}).
For cells with sub-cells, this reduces the \oh{n\log{n}} cost of sorting
to \oh{n} for merging.

\subsection{Task-based implementation}

The particle interactions described in the previous subsection
lead to three basic task types:
\begin{itemize}
    \item Cell {\em sorting}, in which the particles in a given
        cell are sorted with respect to their position along the
        13 cardinal axes, as described in \cite{ref:Gonnet2013}.
    \item Cell {\em self-interaction}, in which all the particles
        of a given cell are interacted with all the other particles
        within the same cell,
    \item Cell {\em pair-interaction}, in which the interactions for
        all particle pairs spanning a pair of cells are computed. 
\end{itemize}
In order to reduce the total number of tasks,
as well as to increase their locality, pair- and self-interactions involving
less than a certain number
of particles are split only implicitly, i.e.~the tasks resulting
from a split are grouped together and executed as a single
task.

The self-interaction and pair-interaction tasks exist in
two flavors, one for the density computation (see \eqn{rho})
and one for the force computation (see \eqn{dvdt}).
Each pair-interaction task requires the sorted indices of
the particles in each cell provided by the sorting tasks.
Since the tasks are restricted to operating on the data of a
single cell, or pair of cells, two tasks conflict if they
operate on overlapping sets of cells.
Due to the hierarchical nature of the spatial decomposition,
two tasks also conflict if any of the cells used by one task
are sub-cells of any of the cells used by the other task.

The interactions have two phases, the~density and force
computation, which need to be clearly separated, i.e.~all the density
tasks on a cell must complete before its force tasks, which
rely on the densities, can be computed.
They are therefore separated by a {\em ghost}
task for each cell.
This ghost task depends on all the density computations
for a given cell, and, in turn, all force computations involving
that cell depend on its ghost task.
This mechanism enforces that all density computations
for a set of particles have completed before this
density is used in any force computations.

Finally, an integrator task for each cell is used to update the
particle positions, velocities, and internal energy once the force
equations of the particles therein have been evaluated.

\begin{figure}
    \centerline{\epsfig{file=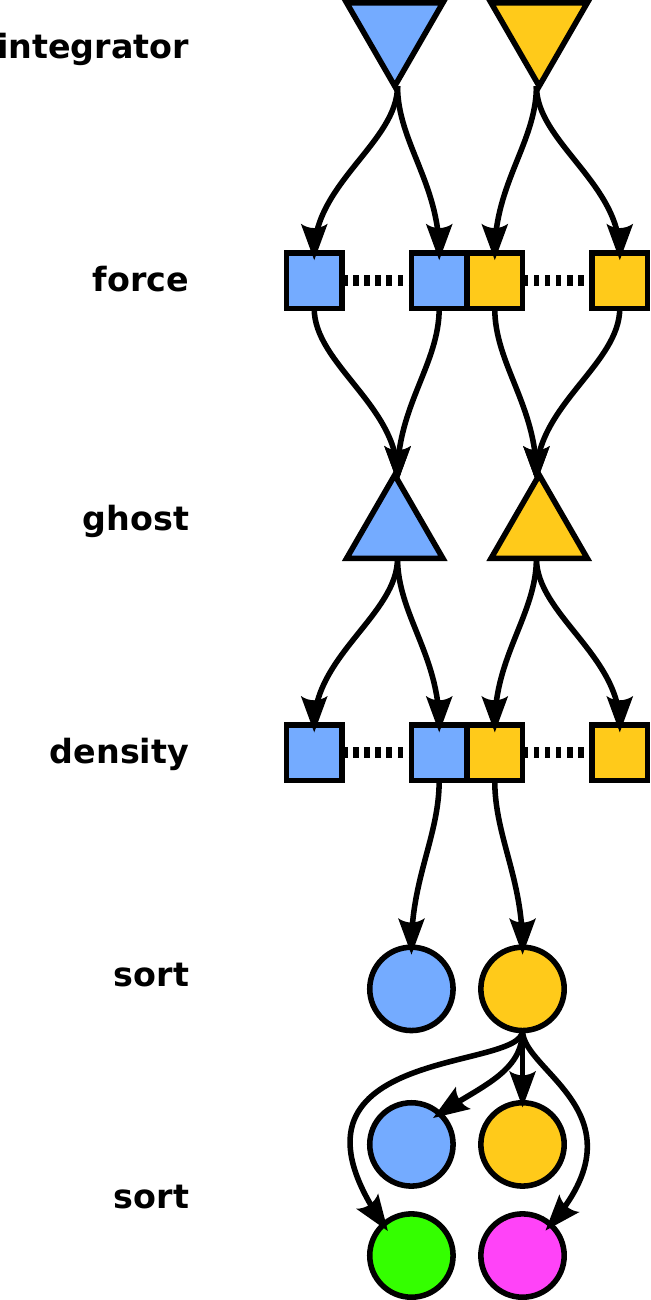,height=0.5\textwidth}}
    
    \caption{Task dependencies and conflicts:
        Arrows indicate the dependencies
        between different task types, i.e.~and arrow from task A to task
        B indicates that A depends on B.
        Dashed lines between tasks indicate conflicts, i.e.~the two tasks
        can not be executed concurrently.
        Each sort task (circles) depends
        only on the sort tasks of its sub-cells.
        The pair-interactions (rectangles) for the particle
        density computation depend on the sort tasks of the respective cells,
        whereas self-interaction tasks (squares) for the density computation
        do not, as they do not require sorting.
        Self- and pair-Interactions on overlapping cells (same color)
        conflict with each other.
        The ghost task of each cell (triangles) depends on the self-
        and pair-interaction density tasks.
        The self- and pair-interaction tasks for the force computation,
        depend on the ghost tasks of the respective cells,
        and the integrator task (inverted triangle) of each cell depends on the cell's
        force tasks.
        }
    \label{fig:Hierarchy2}
\end{figure}

The dependencies and conflicts between the different task types,
which are illustrated in \fig{Hierarchy2},
can be formulated as follows:

\begin{itemize}

    \item Each cell sorting task on a cell with sub-cells depends
        on the sorting tasks of all its sub-cells.

    \item Each cell pair-interaction task depends on the cell sorting
        tasks of both its cells.
        
    \item Cell pair-interaction and cell self-interaction tasks
        operating on overlapping sets of cells or sub-cells
        conflict with each other.
        
    \item The ghost task of each cell depends on all the density cell pair
        interactions and self-interactions which involve the particles
        in that cell.
        
    \item The ghost task of each cell depends on the ghost tasks of
        its sub-cells.
        
    \item Each force cell pair-interaction or self-interaction task
        depends on the ghost tasks of the cells on which it operates.
        
    \item Each integrator task depends on the force cell pair-interaction
        and self-interaction tasks of its cell and of all its cell's
        sub-cells.

\end{itemize}

This task decomposition has significant advantages over the use of
spatial trees.
First of all, the cost of identifying all particles in range of
a given particle does not depend on the total number of
particles, but only on the local particle density.
Furthermore, the particle interactions in each task are computed
symmetrically, i.e.~each particle pair is identified only
once for each interaction type.
The sorted particle indices can be re-used for both the
density and force computation, and even over several time-steps
\cite{ref:Gonnet2013}, thus reducing the computational cost even
further.
Finally, if the particles are stored grouped by cell, each task
then only involves accessing and modifying a limited
and contiguous region of memory, thus greatly improving
cache re-use \cite{ref:Fomin2011}.

\subsubsection{Task scheduling}

The assignment of individual tasks to the processors of a system
is a tricky issue:
Each task must be scheduled once all its dependencies are met,
only if it has no conflicts, and only to a single processor.
Additionally, the tasks should be scheduled in a way that
maximizes both the amount of non-conflicting tasks available to all
processors, as well as the amount of cache re-use between tasks.

In the current implementation, tasks are represented as follows:
        
\begin{center}\begin{minipage}{0.8\textwidth}
    \begin{lstlisting}
struct task {
  int type;
  int wait;
  int nr_unlocks;
  struct task *unlocks;
  ...
};
    \end{lstlisting}
\end{minipage}\end{center}

\noindent where {\tt type} is determines the task type, e.g.~sorting,
pair, or ghost tasks.
The variable {\tt wait} is a counter for the number of
unresolved dependencies belonging to this task and will be
zero when this task is ready to run.
Conversely, {\tt unlocks} is an array of {\tt nr\_unlocks}
pointers to tasks that depend on this task.
The tasks also contain additional data specific to each task
type, i.e.~the cells on which it operates.

The {\tt wait} counters, which are initialized to zero, are
set before the computation by traversing the list of tasks, and,
for each task, incrementing by one the {\tt wait} counter of each of its
{\tt unlocks} tasks.
The {\tt wait} counter then holds the number of other tasks on which
the task depends.
Once a task has been executed, the {\tt wait} counters of its
{\tt unlocks} tasks are decremented by one.
If a task's {\tt wait} counter is zero, then it has no
unsatisfied dependencies and can be executed.

As opposed to the dependencies, the task conflicts are not
encoded explicitly, but implemented via locks on shared resources,
i.e.~the cells on which a task operates.
This locking is described in detail in \sect{locking}.

The tasks are managed by a {\em scheduler}, which assigns tasks
whose dependencies have all been met to different {\em queues}.
The processors, or threads, in a system then obtain the tasks
directly from this scheduler, which tries to obtain a task
from the thread's preferred queue.
If this queue is empty or has no conflict-free tasks, the scheduler
attempts to obtain a task from any other non-empty queue.
This is, in essence, the well-known concept of {\em work-stealing}
described in \cite{ref:Blumofe1999}.

In order to preserve memory locality and improve cache re-use,
the scheduler attempts to assign tasks working on similar sets
of cells to the same queue.
This is done by assigning each cell a pre-defined preferred queue.
Tasks involving only cells of a given queue are assigned to that
queue, and tasks involving more than one queue are assigned to
the shortest of the set of queues.

While the scheduler is responsible for dependencies and data
locality, the queues themselves are responsible for conflict
avoidance and task order.
The queues are implemented as binary heaps which order the tasks
according to their {\em weight}.
A task's weight is defined as the approximate or
measured computational cost of the task, plus the maximum
weight of all its dependent tasks.
This weight is a measure for the length of the critical
execution path starting at the given task.
Picking the task with the largest weight corresponds to reducing
the longest critical path of the task DAG first.

When a processor requests a task from a queue, the heap is
traversed in topological order\footnote{
This is implemented by storing the heap nodes in an array such that the $k$th
entry has sub-nodes at the indices $2k+1$ and $2k+2$, and traversing
this array from left to right.}, starting from the top, looking
for tasks free of conflicts.
Although the first task inspected will have the largest weight
in the queue, the traversal is not in strictly decreasing weight
order.
This sub-optimal traversal was chosen as a efficiency trade-off,
since the cost of inserting or removing an element in a binary heap is in
\oh{\log n}, whereas maintaining a strict ordering requires
at least \oh{n} for insertion, e.g.~using linked lists.
In order to ensure that each task is assigned only once, mutexes
are used to control exclusive access to each queue.

The details of the task scheduler are further described in a separate
publication \cite{ref:Gonnet2013b}.

\subsubsection{Cell locking}
\label{sec:locking}

Particles within a cell are also within that cell's hierarchical
parents.
Therefore, when working on the particles of a cell, tasks which
operate on its parent's data should not be allowed to execute.
One way to avoid this problem is to require that a task
not only lock a cell, but also all of its hierarchical
parents in order to operate on its data.
This, however, would prevent tasks involving siblings who share
a common hierarchical parent cell, yet
whose particle sets do not overlap, from executing.

This problem is avoided by giving each cell both a {\em lock},
and a {\em hold} counter:
A cell is {\em locked} when it, or one of its parent cells, is currently
in use. A cell is {\em held} when one or more of its sub-cells is locked,
and thus cannot be locked itself.
Since more than one task at a time may hold a cell, this property
is implemented as a counter.

The cell locking/holding is implemented as follows:
        
\begin{center}\begin{minipage}{0.8\textwidth}
    \begin{lstlisting}
int cell_locktree(struct cell c) {
  struct cell *c1, *c2;
  if (trylock(c->lock) != 0) return 1;
  if (c->hold > 0) {
    unlock(c->lock)
    return 1;
  }
  for (c1 = c->parent; c1 != NULL; c1 = c1->parent) {
    if (trylock(c1->lock) != 0) break;
    atomic_add(c1->hold, 1);
    unlock(c1->lock);
  }
  if (finger != NULL) {
    for (c2 = c->parent; c2 != c1; c2 = c2->parent) {
      atomic_sub(c2->hold, 1);
    }
    unlock(c->lock);
    return 1;
  } else {
    return 0;
  }
}
    \end{lstlisting}
\end{minipage}\end{center}

\begin{figure}
    \centerline{\epsfig{file=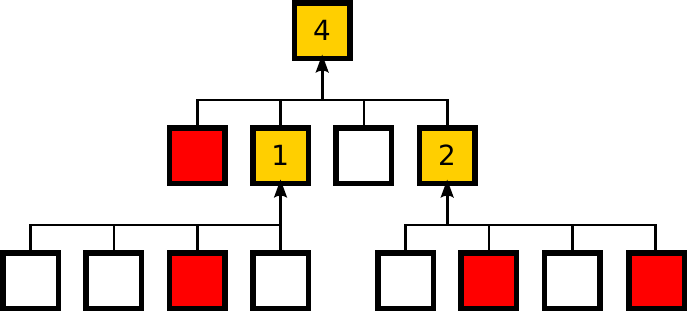,width=0.5\textwidth}}
    
    \caption{Example of hierarchical cell locking. The cells marked in red
        are ``locked'' while the cells marked in yellow have a ``hold'' count
        larger than zero.
        The hold count is shown inside each cell and corresponds to the number
        of locked cells hierarchically below it.
        All cells except for those locked, below a locked cell, or with
        a ``hold'' count larger than
        zero can still be locked without violating any constraints.
        }
    \label{fig:CellLocking}
\end{figure}

\noindent When trying to lock a cell, the function first checks that it is neither
locked (line 3) or held (line 4), i.e.~its hold counter is zero.
If neither is the case, then the cell can be locked.
It then travels up the hierarchy increasing the 
hold counter of each cell on the way, up to the topmost cell (lines 8--12).
If any cell along the hierarchy is locked (line 9), the locking is aborted
and all locks and holds are undone (lines 13--18, see \fig{CellLocking}).
The operations {\tt atomic\_add} and {\tt atomic\_sub} are understood,
respectively, to increase or decrease a value atomically.

When the cell is released, its lock is unlocked and the hold
counter of all hierarchical parents is decreased by one:

\begin{center}\begin{minipage}{0.8\textwidth}
    \begin{lstlisting}
void cell_unlocktree ( struct cell c ) {
    struct cell *c1;
    unlock( c->lock )
    for ( c1 = c->parent ; c1 != NULL ; c1 = c1->parent )
        atomic_sub( c1->hold , 1 );
    }
    \end{lstlisting}
\end{minipage}\end{center}

\subsection{Hybrid shared/distributed-memory parallelism}

Although the task-based algorithms described in the previous subsections
were described only in the context of shared-memory parallelism,
the task-based scheme
extends rather elegantly to {\em hybrid shared/distributed-memory
parallel} setups as well.

The highest-level cells in the hierarchical cell lists can be distributed
between a set of distributed-memory {\em nodes}, such that for each
node the space is partitioned into {\em local} and {\em foreign} cells.
On each node, the tasks are constructed as in the single-node
case, yet each node only keeps the tasks which involve at
least one local cell, i.e. self interaction, ghost, and integrator
tasks on local cells, and pair interaction tasks involving two
local cells, or a local and a foreign cell.
This step would seem to imply that each node would have to have
information on all particles in the system, but it is actually
sufficient for it to know only the structure of the cell hierarchy
of the highest-level foreign cells adjacent to a local cell.

The set of foreign cells which are involved in a pair interaction
with a local cell will be referred to as {\em proxy} cells.
These cells are used to compute interactions locally, but contain
particles which reside on a different node.

\begin{figure}
    \centerline{\epsfig{file=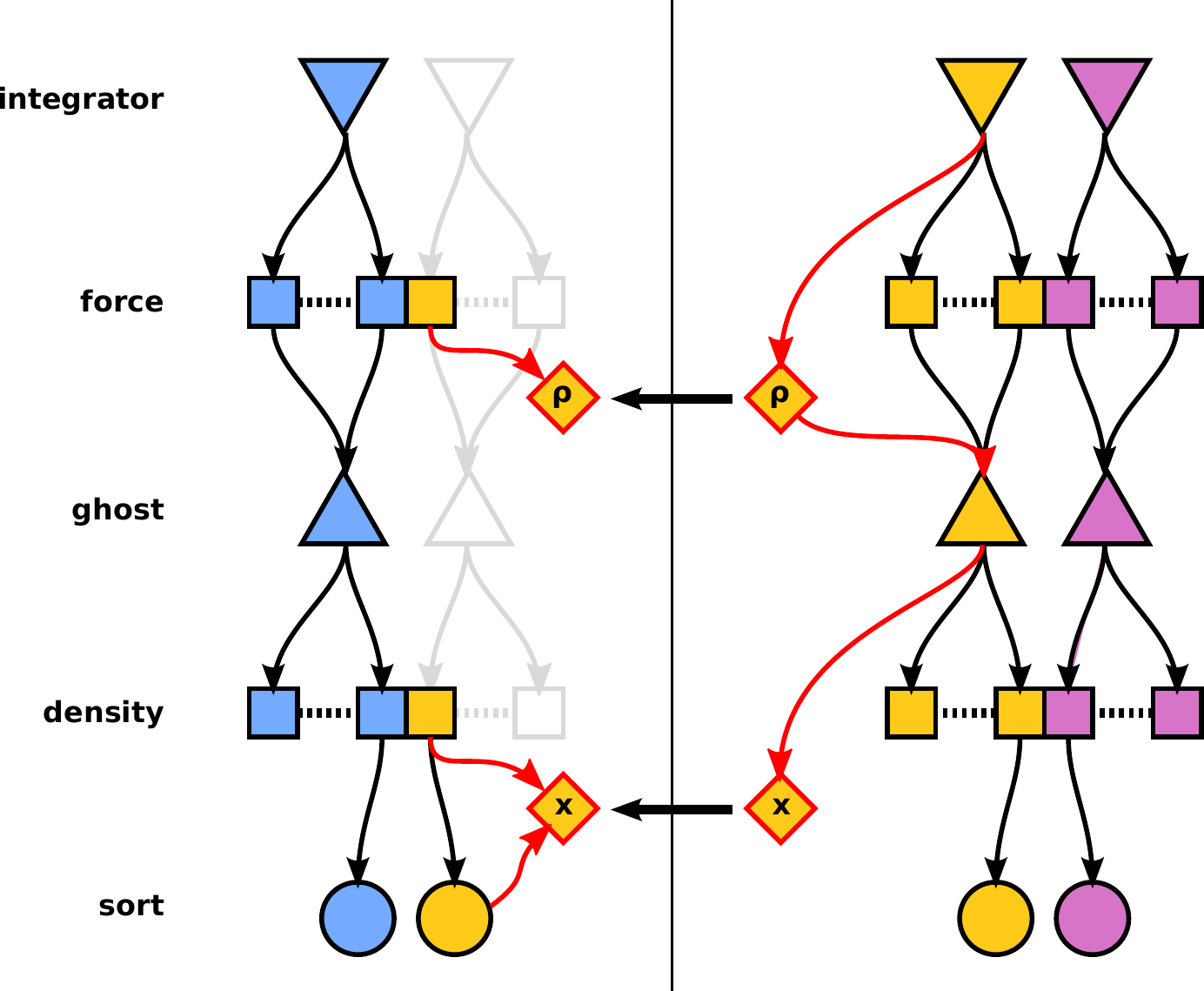,height=0.5\textwidth}}
    
    \caption{Task dependencies and conflicts for hybrid parallelism:
        Two task hierarchies on neighbouring nodes.
        The blue cell, which resides on the left node, interacts
        with the yellow cell, which resides on the right node.
        The regular task hierarchy of the ``ghost'' yellow cell on the
        left is replaced with communication tasks (yellow diamonds)
        and new dependencies (red arrows) on both nodes.
        On the left node, the sorting and density tasks for the yellow cell
        depend on the receipt of the particle positions (marked ``{\bf x}'').
        Similarly, the force tasks depend on the receipt of the
        particle densities (marked ``$\boldsymbol\rho$'').
        On the right node, the sending of the particle densities
        depends on the cell's ghost task to ensure that they have
        been effectively computed.
        The ghost and integrator tasks on the right, which overwrite
        parts of the particle data, are made
        to depend on the communication tasks having completed.
        }
    \label{fig:Hierarchy3}
\end{figure}

For each proxy cell, two {\em communication tasks} are generated, one
to receive the particle positions and other data for the density
computation, and one to receive the particle densities and other
data for the force computation.
The force and density tasks involving each proxy cell are
made dependent of these two communication tasks respectively.
If a proxy cell requires a sort task, the sort task is also made
dependent of the communication task for the particle positions
(see \fig{Hierarchy3}).

Similarly, for each local cell which is a proxy cell on another
node, two communication tasks are generated to send the
particle positions and densities respectively to the foreign node.
The communication task sending a cell's particle positions has
no dependencies, but care must be taken to make any task that
will update the particle data, i.e.~the cell's ghost task, depend
on its completion to avoid corrupting the particle data before
it is sent.
The communication task sending a cell's particle densities depends
on the computation of said densities, i.e.~the cell's ghost task.
Since the cell's integrator task will modify the particle data, it must
be made dependent on the completion of the communication task sending
the cell's particle densities.

All communication is implemented asynchronously using the
{\tt MPI\_Isend}, {\tt MPI\_Irecv}, and {\tt MPI\_Test}
commands.
The {\tt MPI\_Isend} and {\tt MPI\_Irecv} commands for a
communication task are emitted as soon as the task's dependencies
have been met and it is enqueued.
Once in the task queue, the communication tasks are only executed
if a call to {\tt MPI\_Test} returns that they have indeed completed.
The tasks themselves do nothing else, i.e.~all the work happens
in the background or on the calls to {\tt MPI\_Test}, depending
on how asynchronous communication is implemented in the
underlying MPI library.

Contrary to most distributed-memory parallel codes, the
communication between two nodes is not grouped into a single
MPI {\tt send}/{\tt recv} command.
Using individual communication tasks has the advantage that
the particle data, which is stored contiguously for each
cell, does not need to be re-packaged for communication.
It also has the advantage of preventing bottlenecks, since
a single monolithic sending task would have to wait for {\em all} density
tasks involving a different node to complete before it
could be executed.
The only potential disadvantage is that the sum of the latencies
for several small sends and receives is far larger than the
latency for a single large send and receive.
This, however, should be of no particular concern as the task-based
model can just execute other tasks while waiting for the
data to be transmitted, effectively masking any communication latencies.

The main advantage of using non-blocking communication primitives
and a task based model over the traditional
synchronous communication in other distributed-memory parallel codes
is twofold: Firstly, since the computation is data driven,
there are no fixed synchronization points between
nodes, which could force all the nodes to wait
until the slowest node is ready.
Secondly, the communication latencies are completely hidden
i.e.~instead of waiting idly for data to be transmitted, the cores of
a node can execute any other task, if available, in the meantime.

One final advantage associated with using the task-based formulation for
hybrid simulations has to do with the domain decomposition, which
was until here simply assumed to be given.
In most parallel codes, the domain is decomposed following the
simulation {\em data}, i.e.~attempting to spread the particles
evenly amongst the nodes.
The tasks, however, form a complete representation of the {\em work}
that will actually be done during the computation, which can be modelled
as a graph in which each cell is a node, and two nodes share an edge if
they are spanned by a pair-interaction task.
The node weights in the graph are given by the sum of the approximate costs
of all tasks involving only that node, and the edge weights by the
sum of the approximate costs of the tasks spanning both nodes' cells.
This graph can then be partitioned using a standard library
such as METIS \cite{ref:Karypis1998} providing a good equidistribution
of the actual {\em work} across the nodes, as opposed to just
the {\em data}.

\begin{figure}
    \centerline{\epsfig{file=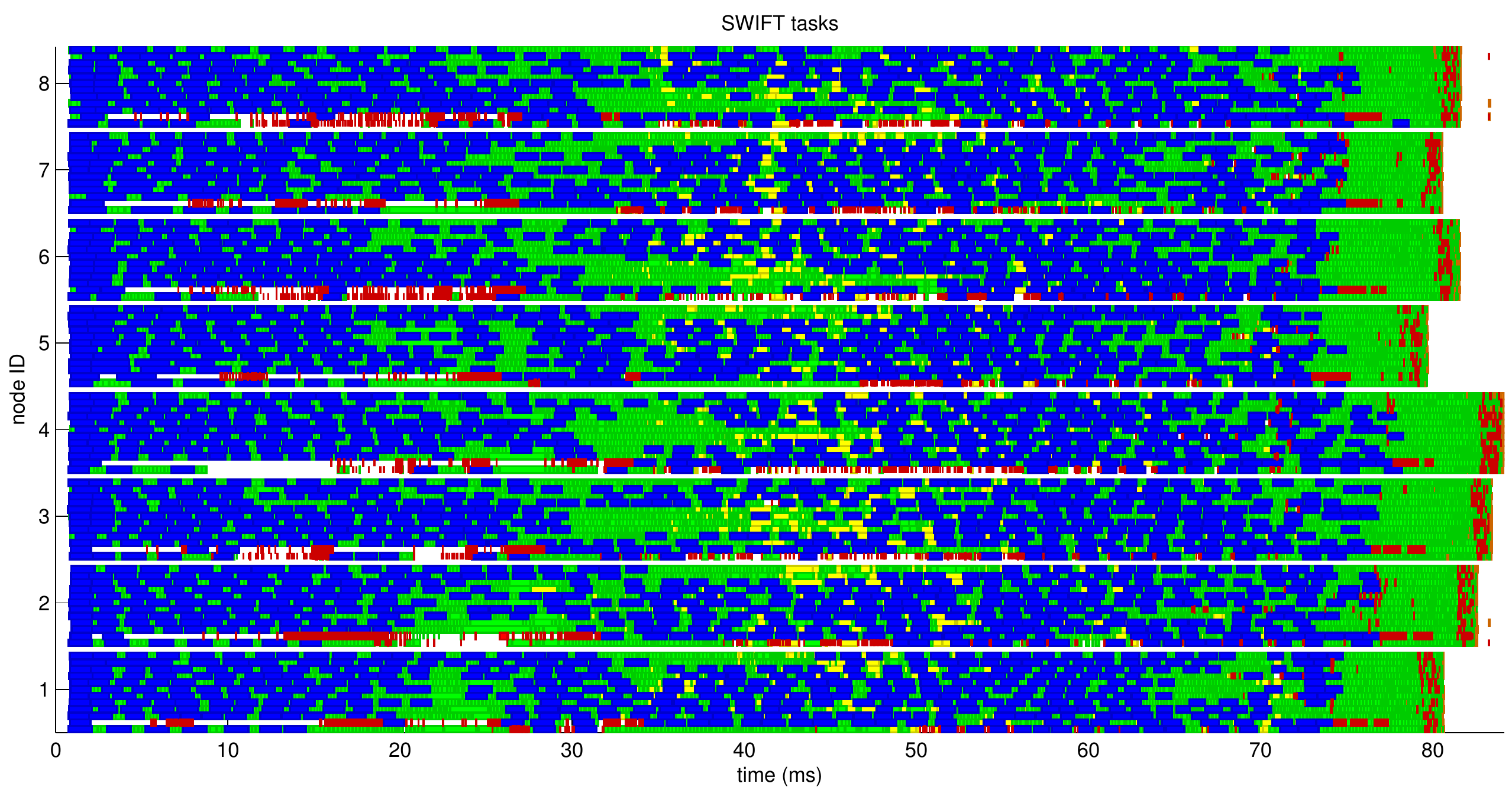,width=0.9\textwidth}}
    \caption{Task timeline for a single time-step of a hybrid
        shared/distributed-memory parallel simulation on
        eight 12-core nodes of the COSMA4 cluster.
        Each task is represented by a block, the color of which represents
        the task type, i.e. pair-interaction (blue), self-interaction (green),
        ghost (yellow), communication (red), and integration (orange).}
    \label{fig:TaskPlot}
\end{figure}

\fig{TaskPlot} shows the task timeline for a single time-step of a hybrid
shared/distributed-memory parallel simulation of 1\,M particles on a
perturbed grid on eight 12-core nodes of the
COSMA4\footnote{\url{http://icc.dur.ac.uk/index.php?content=Computing/Cosma}} cluster.
The different colors indicate the different task types.
The white gaps preceding the red communication tasks are time spent
in the {\tt MPI\_Test} function.
Note that despite the large degree of parallelization (96 cores, $<90$\,ms
per time step), the load distribution between nodes is good an the
load balance within each node is almost perfect, despite the uneven
communication costs.
        
It should be noted that the hybrid approach will still suffer from the
same surface-to-volume ratio problem for distributed computations
described in the introduction.
However, as opposed to purely distributed-memory parallel based
codes, the problem is somewhat mitigated by the fact that
the hybrid approach uses one MPI node per {\em physical node},
as opposed to one MPI node per {\em core}, and that the commuication
latencies can be overlapped with computations over local cells.
With the increasing number of cores per cluster node,
this significantly reduces the ratio of communication per
computation.

\section{Validation}

This section describes how the algorithms shown in the previous section
are implemented and tested against exiting codes on specific problems.

\subsection{Implementation details}

The algorithms described above are all implemented as part
of SWIFT (\underline{S}PH \underline{W}ith
\underline{I}nter-dependent \underline{F}ine-grained
\underline{T}asking),
an Open-Source platform for hybrid shared/distributed-memory
SPH simulations\footnote{See http://swiftsim.sourceforge.net/}.
The code is being developed in collaboration with the Institute
of Computational Cosmology (ICC) at Durham University.

SWIFT is implemented in C, and can be compiled with the
{\tt gcc} compiler.
Although explicitly SIMD-vectorized code, using the {\tt gcc} vector types
and SSE/AVX intrinsics, has been implemented, it was switched
off in the following to allow for a fair comparison with
Gadget-2, which does not use explicit vectorization.

The underlying multithreading for the task-based parallelism is
implemented using standard {\tt pthread}s \cite{ref:pthreads}.
Each thread is assigned it's own task queue.
The threads executing the tasks are initialized once at the
start of the simulation
and synchronize via a barrier between time steps.
The task selection is implemented as described above, yet with the
addition that the ID of the last thread to have worked on each
cell is recorded and the queue inspects up to 50 valid tasks 
looking for one that involves previously-used cells.

The equations of motion are integrated using a velocity-Verlet
integrator.
Multiple time-stepping has been implemented similarly to
the Gadget-2 code \cite{ref:Springel2005}: The maximum time-step
for each particle is computed as
\begin{equation}
    \Delta t_i = C_{CFL}\frac{2 h_i}{ \max_j\left( c_i + c_j + \max\{0,-3 \mathbf r_{ij} \cdot \mathbf v_{ij} / r_{ij} \} \right) }
    \label{eqn:dt}
\end{equation}
where $\mathbf v_{ij} = \mathbf v_i - \mathbf v_j$,
$\mathbf r_{ij} = \mathbf r_i - \mathbf r_j$, and $c_i$ and $c_j$
are the speed of sound of the respective interacting particles.
The CFL constant $C_{CFL}$ is usually set to $\sim 0.3$.
Given a base time-step $\Delta t$, particles for which
$2^{k-1}\Delta t < \Delta t_i \leq 2^k\Delta t$ are only {\em active},
i.e. included in the density and force calculations, every $2^k$th step.
Tasks which do not involve any cell with active particles, and
are not dependencies of any tasks with active particles, are
omitted from the task list in each step.

SWIFT also implements the pseudo-Verlet lists described in
\cite{ref:Gonnet2013}: The spatial decomposition is computed once
and then used over several time-steps until they are invalidated
by particle movement.
The sorted particle indices for each cell are computed and stored
whenever the cells are updated, and re-used over subsequent time-steps.
If the cell decomposition does not need to be re-computed often,
this can lead to substantial savings by eliminating the sort tasks
in these time-steps.

Finally, SWIFT also implements an artificial viscosity of the
Monaghan--Balsara type \cite{ref:Monaghan1983,ref:Balsara1995},
i.e.~the terms
\begin{eqnarray*}
    \frac{dv_i}{dt} & = & -\frac{1}{4}\sum_{r_{ij} < \hat{h}_{ij}} m_j \Pi_{ij} \left({\nabla_r}W({r}_{ij},
    h_i)+{\nabla_r}W({r}_{ij}, h_j)\right) (f_i+f_j),\\
    \frac{du_i}{dt} & = & \frac{1}{8} \sum_{r_{ij} < \hat{h}_{ij}} m_j \Pi_{ij}(\mathbf{v}_i - \mathbf{v}_j)
    \left({\nabla_r}W({r}_{ij},
    h_i)+{\nabla_r}W({r}_{ij}, h_j)\right) (f_i+f_j),
\end{eqnarray*}
are added to \eqn{dvdt} and \eqn{dudt} respectively, where
\begin{eqnarray*}
    \Pi_{ij} &=& -\alpha \frac{\left(c_i + c_j - 3w_{ij}\right)w_{ij}}{\rho_i + \rho_j}, \\
    w_{ij} &=& \min\left\{0, \mathbf{v}_{ij}\cdot\mathbf{r}_{ij} / r_{ij}\right\}, \\
    f_i &=& \frac{|\nabla \times \mathbf{v}_i|}{|\nabla \cdot \mathbf{v}_i| + |\nabla \times \mathbf{v}_i| +
    10^{-4}\frac{c_j}{h_j}}, \\
    \nabla \times \mathbf{v}_i &=& -\frac{1}{\rho_i}\sum_j m_j (\mathbf{v}_j - \mathbf{v}_i)\times
    {\nabla_r}W({r}_{ij}, h_i), \\
    \nabla \cdot \mathbf{v}_i &=& \frac{1}{\rho_i}\sum_j m_j (\mathbf{v}_j - \mathbf{v}_i)\cdot {\nabla_r}W({r}_{ij},
    h_i).
\end{eqnarray*}
and the viscosity parameter $\alpha$ is usually chosen in the
range $[0.5,2]$.

\subsection{Simulation setup}

In order to test their accuracy, efficiency, and scaling,
the algorithms described in the previous section
were tested in SWIFT using the following four
simulation setups:
\begin{itemize}
    \item {\em Sod-shock} \cite{ref:Sod1978}: A rectangular periodic
        domain of size $8\times 1 \times 1$ containing a
        high-density region of
        800\,000 particles with $P_i=1$ and $\rho_i=4$ on one half,
        and a low-density region of 200\,000
        particles with $P_i=0.1795$ and $\rho_i=1$ on the other.
        The simulation results can be compared to an analytic
        solution, providing a test case for the accuracy of the
        implementation.
    \item {\em Sedov blast}: A face-centered cubic lattice of
        $101\times 101\times 101$
        particles at rest with $P_i$ and $\rho_i=1$, yet with the
        central 26 particles set to
        $P_i=100$.
        The resulting blast wave provides a good example of
        strong pressure, density, and smoothing length gradients
        for which an analytical solution can be computed.
    \item {\em Cosmological box}: Realistic distribution of matter
        in a periodic volume of universe at redshift $z=0.5$.
        The simulation consists
        of $\sim$51\,M particles with
        a mix of smoothing lengths spanning three orders of magnitude,
        providing a test-case for neighbour finding and parallel
        scaling in a real-world scenario.
        Although cosmological simulations are often run with
        billions of particles \cite{ref:Springel2005}, the number of
        particles used is sufficient for the study of a number of
        interesting phenomenon.
\end{itemize}
In all simulations, the constants $N_{ngb}=48$, $\gamma=5/3$,
$C_{CFL}=1/4$, and $\alpha = 0.8$ were used.
In SWIFT, cells were split if they contained more than 300
particles and more than 87.5\% of the particles had a smoothing
length less than half the cell edge length.
Fr cells or cell pairs containing less than 6000 particles, the
tasks hierarchically below them were grouped into a single task.

The Sod-shock simulation was run both with and without the
sorted particle interaction in order to provide a rough comparison
to traditional neighbour-finding approaches in SPH simulations with
locally constant smoothing lengths.
In all three test cases, results were
computed using a fixed time step and updating the forces on all
particles in each time-step.
In the Sedov blast simulation, the timestep was set to be below
the smallest particle timestep as computed in \eqn{dt}
in each step.

The simulation results compared with Gadget-2 \cite{ref:Springel2005}
in terms of  speed and parallel scaling.
Gadget-2 was compiled with the Intel C Compiler version 2013.0.028
using the options 
\begin{quote}{\tt -DUSE\_IRECV -O3 -ip -fp-model fast -ftz
-no-prec-div -mcmodel=medium}.\end{quote}
SWIFT v.~1.0.0 was compiled with the GNU C Compiler version 4.8
using the options
\begin{quote}{\tt -O3 -ffast-math -fstrict-aliasing
-ftree-vectorize -funroll-loops -mmmx -msse -msse2 -msse3 -mssse3
-msse4.1 -msse4.2 -mavx -fopenmp -march=native -pthread}.\end{quote}
Note that although the compiler switches for the SSE and AVX
vector instruction sets were activated, explicit
SIMD-vectorization, using vector types and/or intrinsics, was not enabled.
For the hybrid shared/distributed memory parallel runs,
Platform~MPI version 9.1.0 was used.

All simulations were run on the COSMA5 supercomputer,\footnote{\url{http://icc.dur.ac.uk/index.php?content=Computing/Cosma}}
consisting of 420 quad-Intel Xeon E5-2670
16-core nodes running at 2.6\,GHz with CentOS release 6.2 Linux
for x86\_64.
The nodes are connected
via Mellanox FDR10 Infiniband in a 2:1 blocking configuration.

\subsection{Results}

\begin{figure}
    \centerline{\epsfig{file=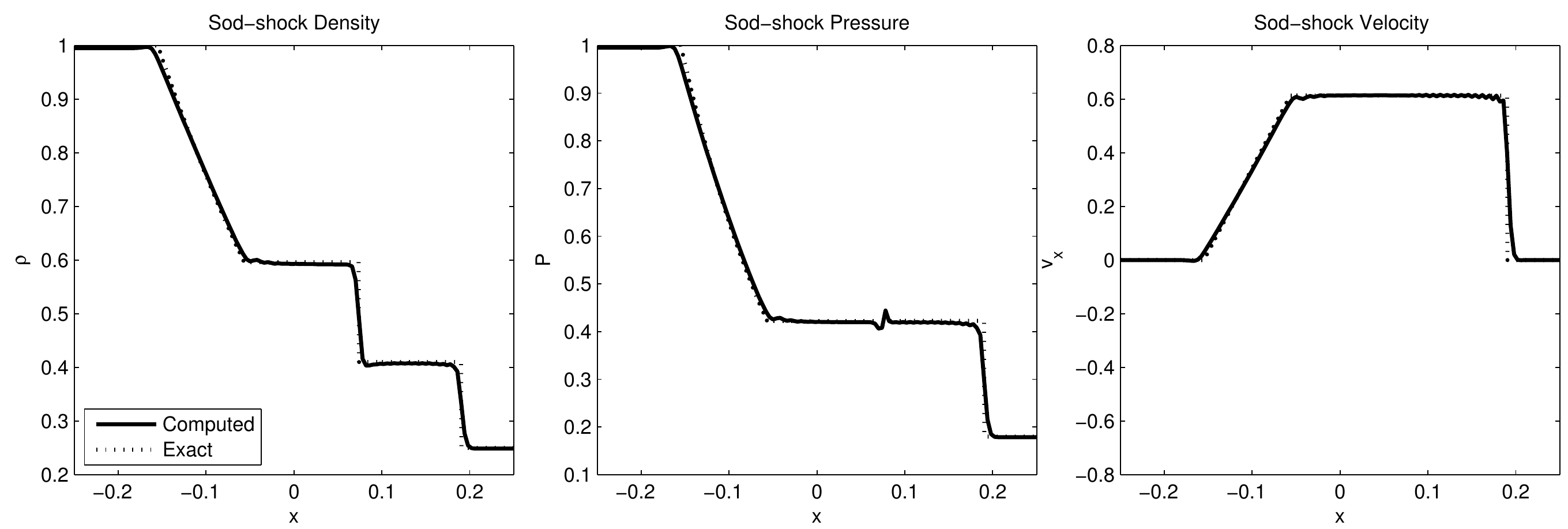,width=0.9\textwidth}}%
    \caption{Results for the Sod-shock simulation test case at $t=0.12$.
        The density,
        pressure, and velocity profiles are in good agreement with the
        analytic solution (top).
        The simulation scales well up to 16 cores of the same shared-memory
        machine, achieving 86\% parallel efficiency (bottom).}
    \label{fig:SodShock}
\end{figure}

\begin{figure}
    \centerline{\epsfig{file=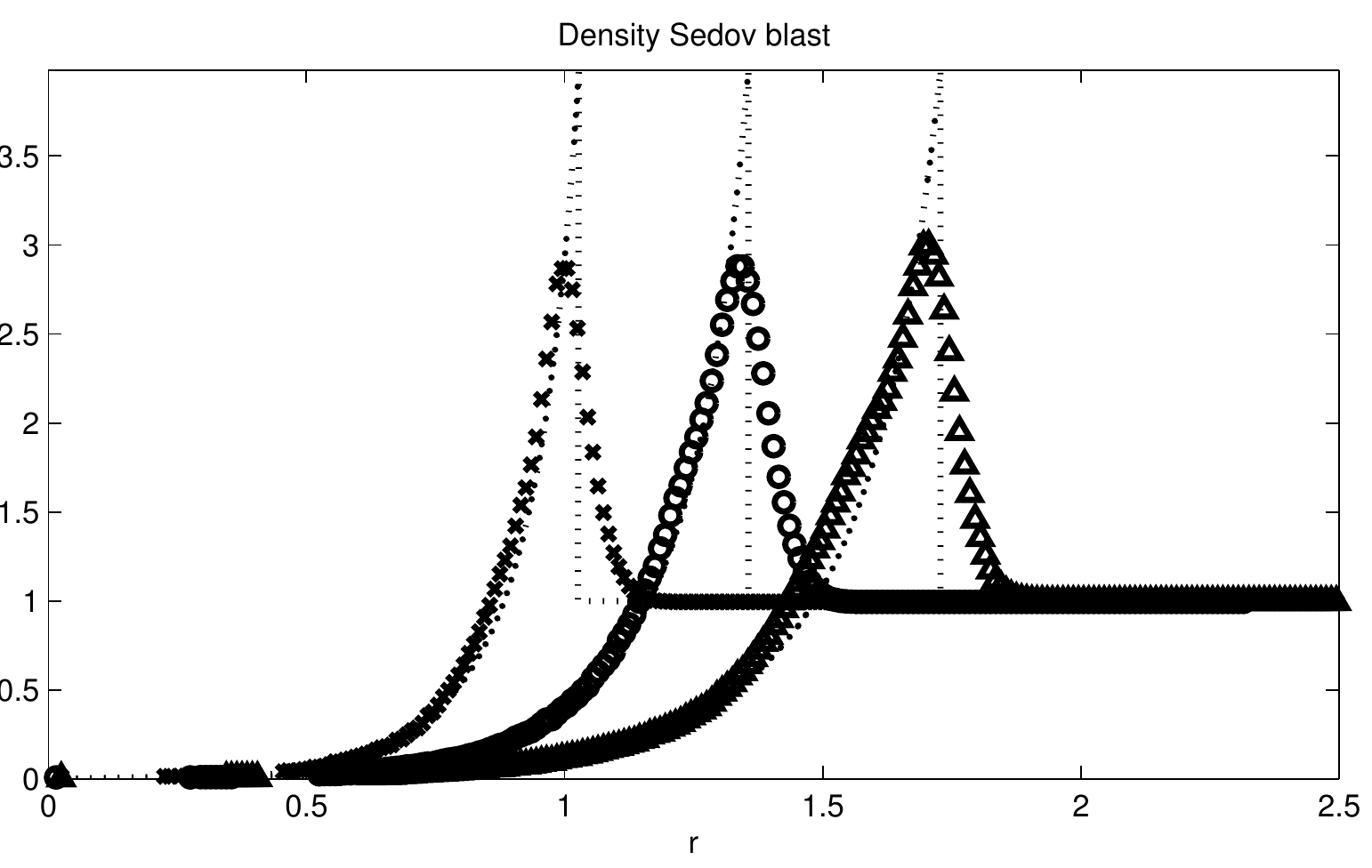,width=0.8\textwidth}}
    \caption{Radial density profile of the Sedov blast simulations at
        times $t=0.075$ (crosses), $t=0.150$ (circles), and $t=0.275$
        (triangles), along with the corresponding analytically computed solution
        (dotted lines).}
    \label{fig:SedovBlast}
\end{figure}

\begin{figure}
    \centerline{\epsfig{file=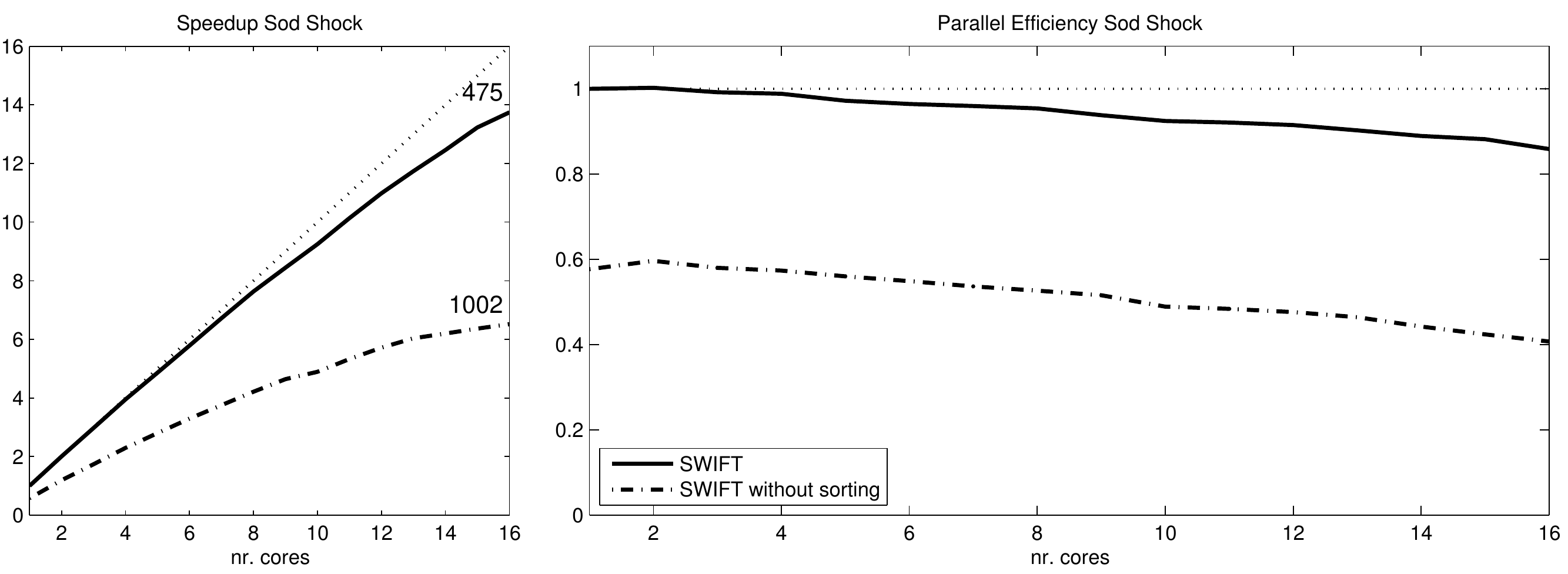,width=0.8\textwidth}}
    \centerline{\epsfig{file=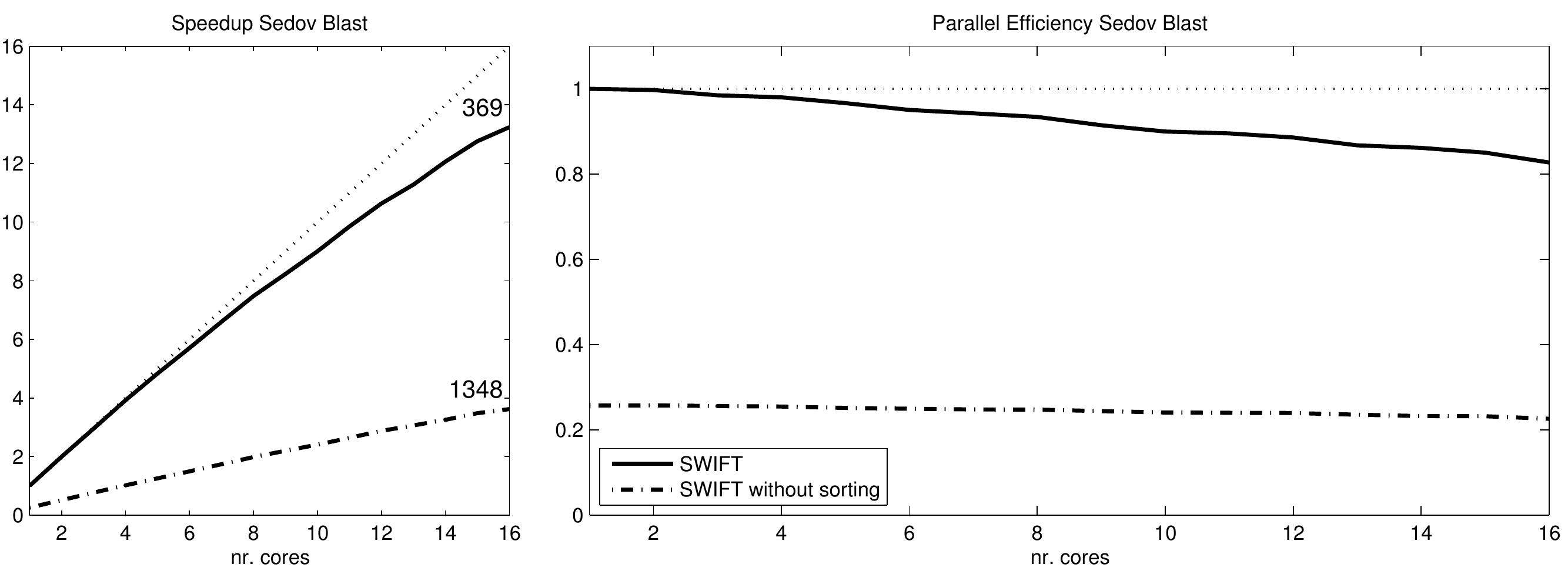,width=0.8\textwidth}}
    \caption{Parallel (strong) scaling and efficiency plots for the Sod-shock and
        Sedov blast simulations on up to 16 cores of a $4\times$Intel Xeon E5-2670
        at 2.6\,GHz. The numbers inside the speedup plots on the left are
        the average number of milliseconds per timestep.}
    \label{fig:Scaling}
\end{figure}

\begin{figure}
    \centerline{\epsfig{file=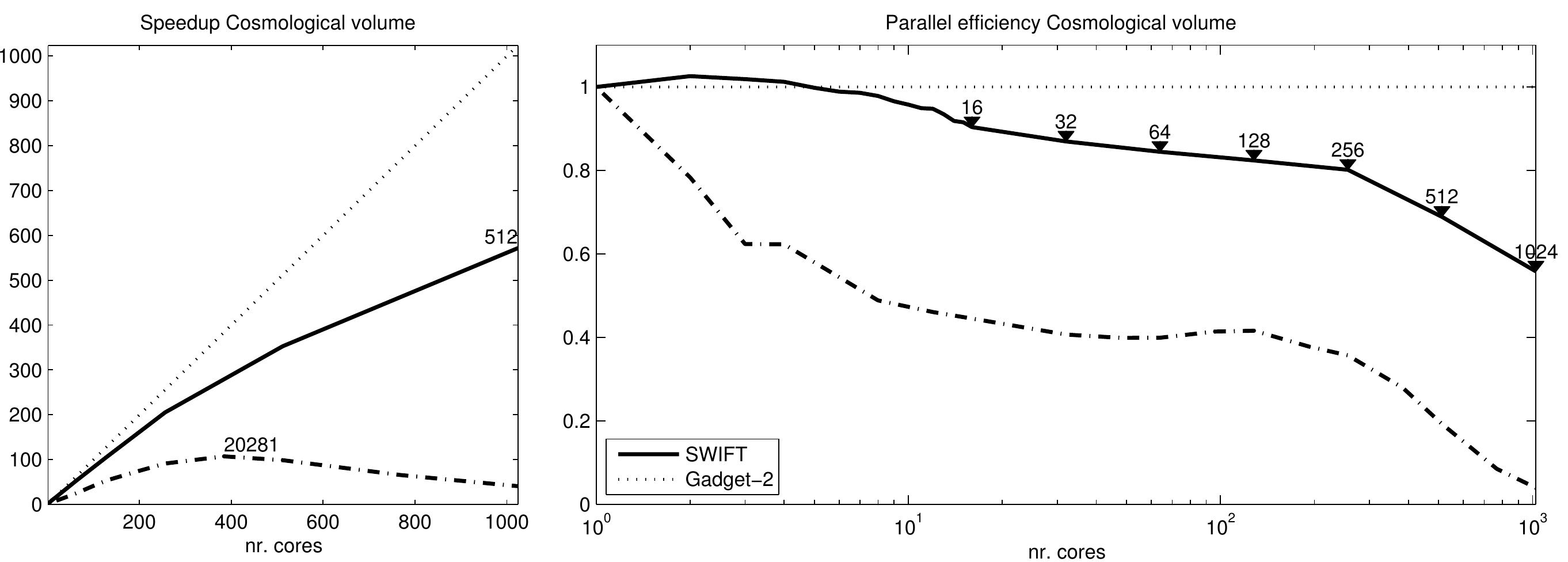,width=0.8\textwidth}}
    \caption{Hybrid parallel (strong) scaling and efficiency for the
        Cosmological volume simulation on up to 64 nodes of the COSMA5
        cluster, containing 16 cores each, for both SWIFT and Gadget-2.
        The numbers in the scaling plot on the left represent the minimum
        average milliseconds per timestep achieved.
        The inverted triangles and numbers in the efficiency plot point to
        the number of cores used at each point.
        For both codes, the parallel efficiency is computed relative to
        the performance of each code on a single core.}
    \label{fig:CosmoVolume}
\end{figure}

\fig{SodShock} shows the averaged density, pressure, and velocity profiles
along the $x$-axis of the Sod-shock simulation at time $t=0.12$.
The computed results are compareable to those produced with Gadget-2 for
the same setup and in good agreement with the analytical solution
\cite{ref:Sod1978}.

Similarly, \fig{SedovBlast} shows the average radial density profiles
for the Sedov blast simulation at times $t=0.075$, $t=0.150$, and $t=0.275$,
along with their analytical solutions \cite{ref:Sedov1959}.

\fig{Scaling} shows the performance of the shared-memory task-based 
code on a single 16-core node of the COSMA5 cluster.
In both cases, SWIFT achieves over 80\% parallel efficiency on all
16 cores.
The Sod-shock and Sedov blast simulations without the sorted
interactions described
in Section~3.3 are roughly $\sim 2-4\times$ slower than the simulation
with sorted interactions.
It should be noted that these results are for {\em strong scaling}
on relatively moderately-sized problems.

The scaling plots are also interesting for what they don't show, namely
any apparent NUMA-related effects, e.g. performance jumps due to
uneven numbers of threads per processor.
Such effects are usually
considered to be a problem for shared-memory parallel codes in
which several cores share the same memory bus and parts of
the cache hierarchy, thus limiting the effective memory bandwidth
and higher-level cache capacities \cite{ref:Thacker2006}.
The hierarchical cells described herein are sufficiently
cache efficient to avoid such problems: By working on sets
of particles which are stored contiguously in memory,
and that fit well in the lower-level processor caches, the ratio
of computation to memory access is kept relatively high, relieving
pressure on the higher-level caches and the memory bus.

Finally, \fig{CosmoVolume} shows the results of the hybrid
shared/distributed-memory parallel Cosmological volume simulation.
On 16 cores of the same node, SWIFT achieves 90\% parallel efficiency.
As the number of nodes is successively doubled, the parallel
efficiency stays above 80\% up to 256 cores (16 nodes), and
drops down to 56\% at 1024 cores (64 nodes), at a speedup
of a factor of 571$\times$.
Gadget-2 does not fare as well, achieving only 40\% parallel efficiency
on a single 16-core node and reaching its maximum speedup of 107$\times$ at
384 cores (24 nodes) with a parallel efficiency of only 28\%.
As of 384 cores, the parallel efficiency decays rapidly, reaching
4\% at 1024 cores.
Over all 1024 cores, SWIFT is 39.6$\times$ faster than Gadget-2.
This speedup is due not only to better scaling, but also to
the better performance on a single core, where SWIFT is already
7.5$\times$ faster than Gadget-2.
Note again that all these results refer to {\em strong scaling},
and that SWIFT was run without explicit vectorization in order
to provide a fair comparison.

%
%

\section{Conclusions}

The good results presented in the previous section can be attributed to
a combination of several factors:

\begin{itemize}

    \item {\em Better algorithms}: The cell-based particle interaction
        algorithms described in Section 3 differ significantly from the
        widely-used tree-based neighbour finding approaches.
        The new neighbour finding algorithm's main advantage is
        its amortized cost of \oh{1} operations per particle,
        as opposed to \oh{N^{2/3}} (for a total of $N$ particles)
        for the tree search.
        The sorted interactions add another factor of $2-4\times$ speedup.
        These algorithmic improvements alone make SWIFT a total
        of 7.5$\times$ faster than Gadget-2, as measured with the
        Cosmological volume simulation, on a single core, i.e.~decoupled
        from any improvements to the parallel performance of either code.
        
    \item {\em Better load balancing on a single node}: The main advantage
        of the task-based parallel approach with constraints
        used in SWIFT is that it
        provides automatic fine-grained load balancing.
        This, along with the
        lack of explicit locking, atomics, and/or synchronization,
        results in more than 80\% parallel efficiency
        on a single 16-core node.
        This almost linear scaling translates to a 15-fold advantage
        over Gadget-2 on a single 16-core node for the Cosmological
        volume simulation.
        The small loss of parallel efficiency is due mainly to the remaining
        small serial bits of the code.
        
    \item {\em Better caching behavior}: A subtle additional advantage of
        the task/cell-based particle interaction algorithm is that
        the computation is organized in such a way that it maximizes the
        amount of computation per memory accessed.
        Instead of interacting a single particle with all its neighbours
        at potentially disparate memory locations, the cell-based
        approach computes all interactions between two sets of contiguous
        particles, allowing them to be kept in the lowest level caches
        for each task.
        Furthermore since each task has exclusive access to it's particles'
        data, there is little performance lost to cache coherency maintenance
        across cores.
        This can be seen in the strong scaling up to 16 cores of the same
        machine, and in the total lack of NUMA-related effects.
    
    \item {\em Better load balancing across multiple nodes}: Instead of
        splitting the {\em particles} over a set of distributed-memory nodes,
        SWIFT uses the task graph to partition the {\em work} over
        the nodes, resulting
        in better load balancing and scaling, as shown by the Cosmological
        volume simulation which has a parallel efficiency of 62\% on 64
        nodes, relative to a single node.
        This is all the more impressive considering that there are less than
        800k particles per node and that each time step takes only half
        a second.
    
    \item {\em Hybrid shared/distributed-memory parallelism}: One of the
        main issues with massively parallel codes is that, as the number
        of involved cores increases, so does the ratio of communication
        to computation for each core, resulting in an eventual loss of scaling.
        In taking a hybrid shared/distributed-memory parallel approach,
        the number of communicating nodes, and thus the total communication,
        is reduced by a factor of the number of cores per shared-memory node.
        
    \item {\em Asynchronous distributed-memory communication}: The task-based
        computation lends itself especially well for the implementation
        of asynchronous communication: Each node sends data as soon
        as it is ready, and activates task which depend on communication
        only once the data has been received.
        This avoids any explicit synchronization between nodes to coordinate
        communication, and, by spreading the communication throughout the
        computation, reduces pressure on the communication infrastructure.
        Furthermore, since the idle time between sending and receiving
        data is used for computation, network latencies play a much
        less limiting role for the overall performance.
    
\end{itemize}

Most of these factors are a direct consequence of the task-based
structure of the computation.
This simple yet powerful paradigm is at the core of the results
presented herein.

It should be noted, finally, that the algorithms and performance gains
described herein are not the result of exploiting any single feature
of the short-lived underlying hardware.
All results presented herein were obtained with {\em existing}, commonplace
computer hardware.
The algorithms merely exploit what has been the trend in computer
architecture for the past decade, i.e.~shared-memory parallel multi-core
systems, shared hierarchical memory caches, and limited communication
bandwidth/latency.
These trends still hold true for more modern architectures such as
GPUs and the recent Intel MIC, on which task-based parallelism
is also possible \cite{ref:Chalk2014}.

\section*{Acknowledgments}

The author would like to thank Matthieu Schaller and Tom Theuns of the
Institute for Computational Cosmology (ICC), and Aidan Chalk of the
School of Engineering and Computing Sciences (SECS), at Durham University,
for the ongoing collaboration of which this work is but one of the first
results.

This collaboration would never have happened were it not for Lydia Heck,
also of the ICC at Durham University, who brought the group together 
and also provided access to and expertise 
on the COSMA5 cluster.

This work used the DiRAC Data Centric system at Durham University,
operated by the Institute for Computational Cosmology on behalf of the
STFC DiRAC HPC Facility (www.dirac.ac.uk). This equipment was funded by
BIS National E-infrastructure capital grant ST/K00042X/1, STFC capital
grant ST/H008519/1, and STFC DiRAC Operations grant ST/K003267/1 and
Durham University. DiRAC is part of the National E-Infrastructure.

\nopagebreak
\bibliography{sph}

\end{document}